\documentclass[journal, twoside]{IEEEtran}
\usepackage{indentfirst}
\usepackage{graphicx}
\usepackage{amsmath}
\usepackage{amssymb}
\usepackage{amsfonts}
\usepackage{mathrsfs}
\usepackage{leftidx}
\usepackage{color}
\usepackage{amsmath}
\usepackage{arydshln}
\usepackage{amsthm}
\usepackage{ragged2e}
\usepackage{cite}
\usepackage{enumerate}
\usepackage{longtable}
\usepackage{float}
\usepackage{stfloats}
\usepackage{hyperref}
\usepackage{algpseudocode}
\usepackage{algorithm}
\usepackage[caption=false,font=footnotesize]{subfig}
\usepackage{tabularx}
\usepackage{makecell}
\usepackage{url}
\usepackage{tabularx}

\usepackage{multirow} 
\usepackage{booktabs}
\theoremstyle{plain}

\usepackage{caption}

\newcolumntype{P}[1]{>{\raggedright\arraybackslash\footnotesize}m{#1}}
\newcolumntype{A}[1]{>{\centering\arraybackslash\footnotesize}m{#1}}

\usepackage[table,usenames,dvipsnames]{xcolor}
\definecolor{aa}{RGB}{175,238,238}
\definecolor{bb}{RGB}{255,255,255}
\usepackage{bm}
\usepackage{makecell}

\begin{document}

\title{Generative AI for Physical-Layer Authentication}

\author{Rui Meng,~\IEEEmembership{Member,~IEEE,} Xiqi Cheng, Song Gao, Xiaodong Xu,~\IEEEmembership{Senior Member,~IEEE,} 

Chen Dong, Guoshun Nan,~\IEEEmembership{Member,~IEEE,} 
Xiaofeng Tao,~\IEEEmembership{Senior Member,~IEEE,} 

Ping Zhang,~\IEEEmembership{Fellow,~IEEE,} and Tony Q. S. Quek,~\IEEEmembership{Fellow,~IEEE}

\thanks{
% This work was supported in part by the National Key Research and Development Program of China under Grant 2020YFB1806905; in part by the National Natural Science Foundation of China under Grant 62501066 and under Grant U24B20131; and in part by the Beijing Municipal Natural Science Foundation under Grant L242012. 
% \textit{(Corresponding author: Xiaodong Xu.)}

Rui Meng, Xiqi Cheng, Song Gao, Xiaodong Xu, Chen Dong, and Ping Zhang are with the State Key Laboratory of Networking and Switching Technology, Beijing University of Posts and Telecommunications, Beijing 100876, China (email: buptmengrui@bupt.edu.cn; chengzi@bupt.edu.cn; wkd251292@bupt.edu.cn; xuxiaodong@bupt.edu.cn; dongchen@bupt.edu.cn; pzhang@bupt.edu.cn).

Guoshun Nan and Xiaofeng Tao are with the National Engineering Research Center for Mobile Network Technologies, Beijing University of Posts and Telecommunications, China (e-mail: nanguo2021@bupt.edu.cn; taoxf@bupt.edu.cn).

Tony Q. S. Quek is with the Singapore University of Technology and Design, Singapore (e-mail: tonyquek@sutd.edu.sg).

}}

\maketitle

\begin{abstract}
Recently, Artificial Intelligence (AI)-driven Physical-Layer Authentication (PLA), which focuses on achieving endogenous security and intelligent identity authentication, has attracted considerable interest.
When compared with Discriminative AI (DAI), Generative AI (GAI) offers several advantages, such as fingerprint augmentation, reconstruction, and denoising.
Inspired by these innovations, this paper provides a systematic exploration of GAI's integration with PLA. We commence with a concise review of identity authentication techniques and GAI models. Then, we contrast the limitations of DAI with the potential of GAI in addressing PLA challenges. Specifically, we introduce a structured taxonomy for GAI-enhanced PLA methodologies, encompassing three key stages: fingerprint collection, model training, and performance optimization within the PLA pipeline. Furthermore, we propose a novel PLA framework based on GAI and channel extrapolation for dynamic environments. To demonstrate GAI's efficacy in enhancing PLA robustness, we implement a case study using the Generative Diffusion Model (GDM). Finally, we outline potential future research directions for GAI-based PLA.

\end{abstract}

%文章从引言到结论部分不得超过4500字。不包括图表、标题、摘要和参考文献。
\section{Introduction}
After 5G technology has fully entered the commercialization stage, research on 6G standards has commenced. According to the IMT-2030 framework issued by the International Telecommunication Union (ITU), 6G not only enhances the three original fundamental scenarios of 5G but also introduces three new scenarios: AI and communication, ubiquitous connectivity, and Integrated Sensing and Communication (ISAC). However, as 6G drives the Internet of everything, it may potentially introduce complex identity security threats.

Physical-Layer Authentication (PLA) technology, with its endogenous security mechanisms and lightweight characteristics, has been considered a promising method to compensate for the shortcomings of traditional secure access mechanisms\cite{zeng2022adaptive,meng2025survey,meng2023physical}. It is primarily implemented based on physical-layer fingerprints, including channel fingerprints \cite{meng2023physical,fu2024generative} and radio frequency (RF) fingerprints \cite{guo2024towards,jiang2025radio}. 
Traditionally, PLA is modeled as a hypothesis testing process. In recent years, AI-based PLA has gradually become a research hotspot \cite{lai2025comprehensive}.

Traditional Discriminative AI (DAI) utilizes deep neural networks to learn features from large-scale fingerprint datasets. However, it relies on extensively labeled fingerprints for training. Fortunately, the emergence of Generative AI (GAI) has mitigated these limitations. Instead of learning class boundaries, GAI learns the probability distribution of training fingerprints to generate new fingerprint samples \cite{zhao2025generative}, which is expected to be applied to Internet of Things (IoT) \cite{zeng2022adaptive,liu2025large}, Industrial IoT (IIoT) \cite{meng2023physical}, Internet of Vehicles (IoV) \cite{zhang2021generative}, and satellite communication systems \cite{jiang2025radio}.
For instance, Meng \textit{et al.} \cite{meng2023physical} proposed a Hierarchical Variational Autoencoder (HVAE)-based PLA scheme to detect spoofing attackers at the edge of IIoT. Besides, Liu \textit{et al.} \cite{liu2025large} proposed a Bidirectional Encoder Representation from Transformers (BERT)-based lightweight PLA framework to enhance the zero-trust edge IoT security. Additionally, Jiang \textit{et al.} \cite{jiang2025radio} combined convolutional neural networks, Generative Adversarial Networks (GANs), and Bayesian meta-learning, to realize security authentication of the satellite component in automatic identification systems. Furthermore, Chi \textit{et al.} \cite{chi2024rf} combined Generative Diffusion Models (GDMs) and Transformers to learn the time, frequency and complex-valued domains of RF fingerprints.
Nevertheless, research on GAI-based PLA is still in its early phases. The full potential of applying GAI to PLA has yet to be comprehensively investigated. Motivated by this, we present a systematic investigation into GAI's integration with PLA. The primary contributions are outlined as:
\begin{itemize}
\item We analyze the issues of traditional DAI solved by GAI in PLA, including insufficient fingerprint data, environmental noises and interferences, perturbations in fingerprint data, and complex tasks.
\item We propose a structured taxonomy for GAI-enhanced PLA methods, delineating three stages within the PLA pipeline: fingerprint collection, model training, and performance optimization. The fingerprint collection stage encompasses fingerprint estimation, fingerprint augmentation, fingerprint filling, and fingerprint transmission. The model training stage includes attack detection, multiuser authentication, and efficient training and deployment. The performance optimization stage focuses on the flexibility for eliminated legitimate users,defense against intelligent attacks, and adaptation for different scenarios.
\item We propose a channel extrapolation-based PLA framework integrating GAI, addressing the challenges faced by conventional PLA schemes. We further present a novel case study leveraging GDMs, to show the superiority of GAI in enhancing the robustness of PLA.
\item We outline a future research roadmap for GAI-based PLA across five critical dimensions: high-reliable fingerprints, efficient PLA, generalization enhancement, model security, and industrialization.
\end{itemize}

\section{Overview of Identity Authentication and Generative Artificial Intelligence (GAI)}

\subsection{Identity Authentication}

\subsubsection{Why Identity Authentication?}
Identity authentication serves as a critical component in ensuring wireless identity security, and its necessity is primarily reflected as follows.
\begin{itemize}
\item \textbf{Preventing Unauthorized Access:} 
Identity authentication verifies the authenticity of communicating parties, blocks unauthorized devices at the source, and establishes a trusted communication link \cite{meng2023physical}.
\item \textbf{Safeguarding Data Security:} Identity authentication acts as a trust anchor for encryption systems, ensuring that keys are exclusively held by legitimate parties, thereby preserving the confidentiality and integrity of data transmission \cite{liu2025large}.
\item \textbf{Mitigating Resource Abuse:} Identity authentication can reject devices with forged identities from occupying channel resources or launching Distributed Denial-of-Service (DDoS) attacks, thereby preventing network congestion and enhancing Quality of Service (QoS) \cite{lai2025comprehensive}.
\end{itemize}

\subsubsection{How to Identity Authentication?}
Four representative identity authentication techniques are compared as follows. Cryptography-based upper-layer authentication is mainly used in mobile communication systems, but relies on the limited computation power assumptions of attackers; Blockchain-based authentication is based on a multi-layer framework and a fault-tolerant mechanism, but has high delay and low throughput; Physical-layer-key-based authentication relies on the wireless channel reciprocity, but is difficult to implement in frequency division duplexing systems and static environments. In contrast, PLA leverages physical-layer attributes, including channel fingerprints and RF fingerprints, to realize endogenous security, lightweight, and high compatibility.

\subsection{GAI}
\subsubsection{What is GAI?}
Unlike conventional DAI optimized for classification or regression tasks, GAI excels at learning potential structural patterns within training data, enabling autonomous generation of diverse content such as images, audio, and text. Its strength resides in modeling the distribution of high-dimensional data, facilitating the generation of realistic synthetic datasets, and augmenting sparse training samples.

\subsubsection{How to GAI?} 
Four typical models are presented as follows.

\begin{itemize}
\item \textbf{Generative Adversarial Network (GAN):} 
GANs consist of a generator and a discriminator, participating in an adversarial two-player minimax game. The generator synthesizes data to simulate real samples, while the discriminator evaluates and distinguishes authenticity. Through iterative adversarial training, GANs excel at producing clear and high-fidelity outputs.

\item \textbf{Variational Autoencoder (VAE):} 
VAEs employ probabilistic encoder-decoder architectures to map data into a structured latent space and reconstruct it via probabilistic sampling. The loss function balances the reconstruction accuracy and KL divergence to align the latent distribution to a predefined Gaussian prior.

\item \textbf{Generative Diffusion Model (GDM):} 
GDMs learn to generate data by reversing a progressive noising process. By gradually denoising from random noise, they can model complex distributions with strong diversity and stability. Despite high computational complexity, they provide high-quality and consistent outputs.

\item \textbf{Transformer:} 
Transformers have become the cornerstone of Large Language Models (LLMs) through the unique self-attention mechanism and modular design. For example, ChatGPT combines supervised fine-tuning with reinforcement learning from human feedback to optimize conversational alignment, while DeepSeek incorporates a Mixture of Experts (MoE) architecture and introduces instruction tuning steps during training to improve efficiency and performance. These advancements enable LLMs to handle nuanced tasks, from creative writing to task-oriented dialogue.
\end{itemize}

\begin{table*}[tbp] % 使用默认的[tbp]位置参数
    \centering
    \caption{The outline of the issues faced in the field of PLA, the limitations of DAI in tackling these difficulties, and the promising capabilities of GAI to boost performance within these contexts.}
    \label{tab1}
    \renewcommand{\arraystretch}{1.2}  % 增加行高以改善可读性
    \normalsize  % 设置适中的字体大小
    \begin{tabular}{|>{\arraybackslash}m{0.09\textwidth}|>{\arraybackslash}m{0.42\textwidth}|>{\arraybackslash}m{0.42\textwidth}|}
        \hline
        \textbf{\small{Issues}} & \textbf{\small{Limitations of DAI}} & \textbf{\small{Potential of GAI}} \\
        \hline
        \footnotesize Insufficient fingerprint data  & \footnotesize\hangindent=1em \noindent \textbullet\hspace{0.5em}It relies on a large number of labeled fingerprint samples.
        
        \hangindent=1em \noindent \textbullet\hspace{0.5em}This strict supervised learning mode increases annotation costs.
        
        \hangindent=1em \noindent \textbullet\hspace{0.5em}The potential value of unlabeled fingerprint samples remains underutilized. & \footnotesize\hangindent=1em \noindent \textbullet\hspace{0.5em}It directly learns distributions of fingerprints.
        
        \hangindent=1em \noindent \textbullet\hspace{0.5em}It allows for the controlled generation of fingerprint variants simulating different conditions.

        \hangindent=1em \noindent \textbullet\hspace{0.5em}It fully utilizes the potential value of unlabeled data \cite{zhao2024finding}.\\
        \hline
        \footnotesize Environment noises and inferences & \footnotesize\hangindent=1em \noindent \textbullet\hspace{0.5em}It constructs classification boundaries only under preset SNR conditions.

        \hangindent=1em \noindent \textbullet\hspace{0.5em}It struggles to deeply analyze the complex nonlinear interactions between noise and fingerprints. & \footnotesize\hangindent=1em \noindent \textbullet\hspace{0.5em}It reveals the deep-seated patterns of noise's intrinsic influence on fingerprint features by joint probability modeling \cite{chi2024rf}.

        \hangindent=1em \noindent \textbullet\hspace{0.5em}It breaks the dependence of generated fingerprint samples on specific SNR conditions \cite{zeng2022adaptive}. \\
        \hline
        \footnotesize Perturbations in fingerprint data & \footnotesize\hangindent=1em \noindent \textbullet\hspace{0.5em}It lacks proactive strategies to defend against adversarial attacks.

        \hangindent=1em \noindent \textbullet\hspace{0.5em}It relies on static rules for authentication decisions.
        
        \hangindent=1em \noindent \textbullet\hspace{0.5em}It is highly sensitive to minor changes in input fingerprints. & \footnotesize\hangindent=1em \noindent \textbullet\hspace{0.5em}GAN possesses an inherent defensive advantage due to its adversarial training nature \cite{papangelo2024adversarial}.
        
        \hangindent=1em \noindent \textbullet\hspace{0.5em}It cleverly formulates PLA as probability inference problems.
        
        \hangindent=1em \noindent \textbullet\hspace{0.5em}It can eliminate the impact of perturbation components through the learned fingerprint distribution.
       \\
        \hline
        \footnotesize Complex tasks & \footnotesize\hangindent=1em \noindent \textbullet\hspace{0.5em}When faced with highly similar fingerprint features among multiple users, it is difficult to distinguish them.
        
        \hangindent=1em \noindent \textbullet\hspace{0.5em}It requires retraining to adapt to the updated user categories.
        
        \hangindent=1em \noindent \textbullet\hspace{0.5em}It typically requires separately constructing models for different tasks.  & \footnotesize\hangindent=1em \noindent \textbullet\hspace{0.5em}It builds an exclusive probability distribution model for the fingerprint features of each user.
        
       \hangindent=1em \noindent \textbullet\hspace{0.5em}It introduces fine-tuning methods to adapt to the updated users without retraining the model.
        
        \hangindent=1em \noindent \textbullet\hspace{0.5em}It integrates multi-tasks through joint probability modeling \cite{guo2024towards}.\\
        \hline
    \end{tabular}

\end{table*}

\subsection{What PLA Issues Can Benefit from GAI?}

As depicted in Table \ref{tab1}, the limitations of DAI and GAI's potential in PLA are presented as follows.
\begin{itemize}
\item \textbf{Insufficient Fingerprint Data:}
Given the unpredictable interference caused by environmental complexity and the inherent physical constraints of hardware measurement systems, it is often difficult to collect sufficient high-precision fingerprint samples in actual communication scenarios \cite{meng2023physical}. Traditional DAI methods have limitations in reliance on a large number of fingerprint samples, reliance on upper-layer authentication protocols or manual labeling of fingerprints, and low utilization of unlabeled fingerprint data. In contrast, GAI has potential in augmenting fingerprint data, learning fingerprint data distributions through latent variables, learning unlabeled fingerprint data, and continuously optimizing its model parameters and structure.

\item \textbf{Environment Noises and Inferences:}
In wireless environments, the signal-to-noise ratio (SNR) often fluctuates due to various factors, including changes in path loss caused by mobile device displacement, shadow fading induced by obstacle obstruction, and rapid signal intensity fluctuations resulting from multipath effects \cite{zeng2022adaptive}. Traditional DAI methods have limitations in training PLA models for specific SNR conditions, capturing the coupling relationship between fingerprints and noises or inferences, and requiring prior knowledge of noises. In contrast, GAI has potential in modeling the joint distribution of fingerprints and noises, synthesizing fingerprints under different SNRs through conditional generation, and denoising and reconstructing fingerprints.

\item \textbf{Perturbations in Fingerprint Data:}
Faced with emerging intelligent threats such as adversarial attacks \cite{papangelo2024adversarial}, traditional DAI methods have limitations, lacking defense against well-designed adversarial perturbations, decision making based on fixed rules, and vulnerability and sensitivity to small disturbances. In contrast, GAI has the potential to defend against adversarial attacks through adversarial training, eliminate interference through fingerprint reconstruction, and make flexible inferences, judgments, and decisions based on learned fingerprint probability distributions.

\item \textbf{Complex Tasks:}
For complex tasks such as multiuser authentication and identifying the updated user categories, traditional DAI methods have limitations in fuzzy classification boundaries caused by overlapping fingerprint feature spaces of users and require retraining the PLA model. In contrast, GAI has the potential to generate independent fingerprint distributions for each user to avoid feature overlap and expand to the updated users without retraining the PLA model.
\end{itemize}

\begin{figure*}
\centering
\includegraphics[width=1\textwidth]{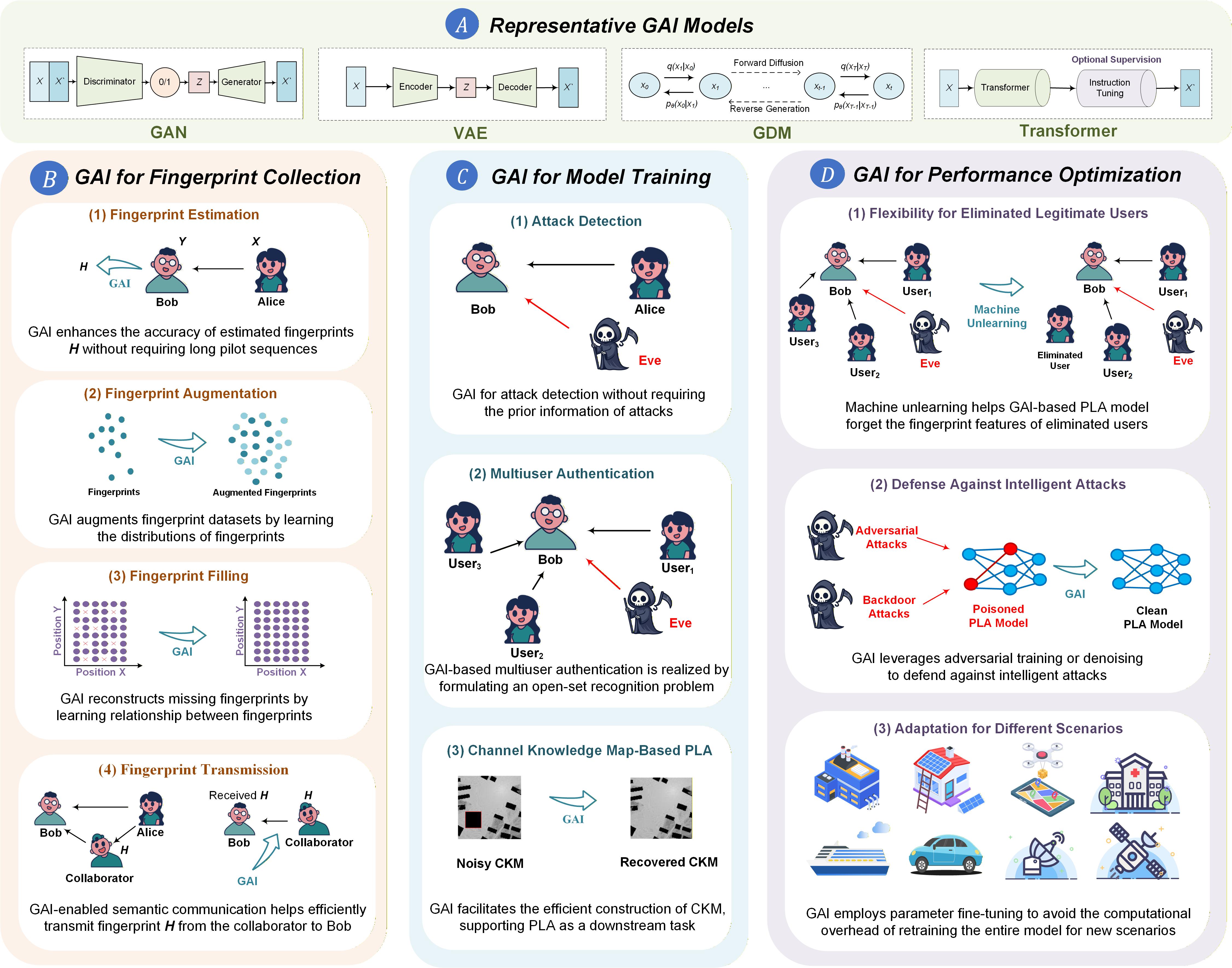}
\caption{Illustration of GAI-enhanced methods for PLA, where \textit{Part (A)} denotes representative GAI models, including Generative Adversarial Network (GAN), Variational Autoencoder (VAE), Generative Diffusion Model (GDM), and Transformer, \textit{Part (B)} denotes GAI for fingerprint collection phase, including fingerprint estimation, fingerprint augmentation, fingerprint filling, and fingerprint transmission, \textit{Part (C)} denotes GAI for PLA model training, including attack detection, multiuser authentication, and channel knowledge map, and \textit{Part (D)} denotes GAI for PLA performance optimization, including flexibility for eliminated legitimate users, defense against intelligent attacks, and adaptation for different scenarios.}
\label{figure1}
\end{figure*}

\begin{table*}[!ht]
    \centering
    \caption{The applications of GAI models in PLA.}
    \label{tab2}
    \renewcommand{\arraystretch}{1.5}  % 调整行高
    \resizebox{\textwidth}{!}{%
    \normalsize  % 设置适中的字体大小
    \begin{tabular}{|p{2cm}|p{3cm}|p{3cm}|p{3cm}|p{3cm}|p{3cm}|}  % 第一列改为 p{2cm}，允许自动换行
    \hline
        \multirow{2}{*}{Phases} & \multirow{2}{*}{Issues} & \multicolumn{4}{c|}{GAI Models} \\ \cline{3-6}
        & ~ & GAN & VAE & GDM & Transformer \\ \hline
        \multirow{4}{*}{\footnotesize \shortstack{Fingerprint\\Collection}} & \footnotesize Fingerprint Estimation & \footnotesize Determining mapping functions of fingerprints through generators \cite{zhang2021generative} &\footnotesize Approximating estimation through variational inference process & \footnotesize Estimating fingerprints through posterior sampling & - \\ \cline{2-6}
        & \footnotesize Fingerprint Augmentation & \footnotesize Conditioning GANs for time-varying channel environments \cite{jiang2025radio} & \footnotesize Decoders for augmenting fingerprints & \footnotesize Generating fingerprint samples during reverse processes \cite{chi2024rf}& - \\ \cline{2-6}
        & \footnotesize Fingerprint Filling & - & - & - & \footnotesize Learning dependencies between fingerprints by attention mechanisms \cite{zhao2024finding} \\ \cline{2-6}
        & \footnotesize Fingerprint Transmission & \footnotesize Reconstructing transmitted signal using generators & \footnotesize Encoders and decoders for joint source-channel coding & \footnotesize Channel adaptive de-noising transmission & \footnotesize Attention-based semantic encoding and decoding  \\ \hline
        \multirow{3}{*}{\footnotesize Model Training} 
        & \footnotesize Attack Detection & \footnotesize Modifying the discriminator's functionality to output category distributions & \footnotesize Latent Space for unsupervised classification \cite{meng2023physical} & \footnotesize Comparing generation probabilities under different identity hypotheses & \footnotesize -\\ \cline{2-6}
        & \footnotesize Multiuser Authentication & \footnotesize Formulating as an OSR problem \cite{guo2024towards} & - & - & - \\ \cline{2-6}
        & \footnotesize Efficient Training and Deployment & \footnotesize Generating diverse fingerprints to meet the requirements of complex environments & \footnotesize Supporting probabilistic modeling and feature decoupling & \footnotesize Formulating as training a conditional denoiser \cite{fu2024generative} & \footnotesize Supporting complex logical reasoning and knowledge transfer \\ \hline

        \multirow{3}{*}{\footnotesize \shortstack{Performance\\Optimization}} 
        & \footnotesize Flexibility for Eliminated Legitimate Users & \footnotesize Gradient-based optimization & \footnotesize Gradient-based optimization & \footnotesize Gradient-based optimization, data sharding, and adding learnable layers & \footnotesize Gradient-based optimization, knowledge distillation, and adding learnable layers \\ \cline{2-6}
        & \footnotesize Defense Against Intelligent Attacks & \footnotesize Adversarial training \cite{papangelo2024adversarial} & \footnotesize Identifying abnormal samples through reconstruction errors & \footnotesize Eliminating fingerprint triggers through denoising process & - \\ \cline{2-6}
        & \footnotesize Adaptation for Different Scenarios & - & - & - & \footnotesize Parameter-efficient fine-tuning techniques \cite{liu2025large} \\ \hline

    \end{tabular}%
    }
\end{table*}

\section{GAI-enhanced Methods for PLA}

As illustrated in Figure \ref{figure1}, we sort out which stages of PLA can be enhanced by GAI, including fingerprint collection, authentication model training, and authentication performance optimization.

\subsection{Fingerprint Collection}
\subsubsection{Fingerprint Estimation}
% Channel estimation is a critical technique for acquiring channel fingerprints. 
DAI-based estimation methods require long pilot sequences to achieve satisfactory performance and struggle in high path loss scenarios. In contrast, GAI can generate synthetic training fingerprints with distributions similar to real training samples, thereby providing effective support for fingerprint estimation. 
% For instance, the variational inference process in VAEs approximates the maximum likelihood process of fingerprint estimation. Similarly, GDMs can learn score functions of posterior fingerprint distributions and employ posterior sampling during reverse processes to recover fingerprint samples. Additionally, 
Zhang \textit{et al.} \cite{zhang2021generative} applied the discriminator in GANs to learn and extract intrinsic features of time-varying channels in high-speed railway systems, while the generator determines implicit mapping functions of training fingerprint data to enhance estimation performance.

\subsubsection{Fingerprint Augmentation}

GAI can augment fingerprint datasets, thereby enhancing the robustness of PLA models. For instance, Jiang \textit{et al.} \cite{jiang2025radio} employed conditional GANs to synthesize fingerprint samples for Low Earth Orbit (LEO) satellites. 
% By conditioning GANs on time-varying channel environments, they addressed a critical limitation of conventional GANs, the lack of control over generated fingerprints. 
This conditional approach enables the synthesis of more realistic fingerprints, as environmental factors are explicitly incorporated into the generation process.
% Similarly, Chi \textit{et al.} \cite{chi2024rf} employed GDMs to jointly leverage temporal, spectral, and complex-valued features of fingerprint waveforms, thus enhancing the generation performance of fingerprints. 
% Specifically, they developed a hierarchical diffusion Transformer, a modular neural architecture that operationalizes the theory through strategic design choices in network topology, functional components, and complex-number operations.

\subsubsection{Fingerprint Filling}

In the real-world, fingerprint data loss may occur due to physical obstructions, device malfunctions, and environmental disturbances. Such missing fingerprint data directly disrupts the spatiotemporal continuity of fingerprint sequences, necessitating fingerprint filling methods to restore complete fingerprint sequences.
DAI-based methods establish data correlations only between adjacent nodes through local neighborhood modeling, making them ill-suited to adapt to dynamic changes in node positions or obstruction patterns.
In contrast, GAI models, such as Transformers, leverage self-attention mechanisms to automatically learn dependency strengths between different fingerprint data points, dynamically capturing long-range spatiotemporal correlations, thereby achieving more robust missing fingerprint reconstruction performance \cite{zhao2024finding}.
% For example, Zhao \textit{et al.} \cite{zhao2024finding} introduced a Bidirectional Encoder Representations from Transformers (BERT)-based framework for robust fingerprint filling. 

% Trained in a self-supervised manner on target datasets without external data requirements, the proposed scheme distinguishes itself from conventional interpolation methods by jointly modeling sequential dependencies across multiple subcarriers, rather than processing each independently.

\subsubsection{Fingerprint Transmission}
% Collaborative PLA frameworks, leveraging multi-party cooperation, demonstrate superior authentication efficacy compared to centralized approaches. 
% % This advantage stems from the deployment of cooperative nodes across diverse geographic locations, which harnesses spatial diversity to bolster the robustness of fingerprints.
% However, a critical challenge lies in efficiently transmitting collaborator-obtained fingerprints to Bob.
% As the dimensionality of CSI samples grows, so too does the requirement for a larger number of such samples to ensure robust authentication performance, further compounding the challenge of the fingerprint transmission efficiency. 
% Conventional communication systems have approached theoretical performance limits under classical information theory. In contrast, GAI-enabled semantic communication (SemCom) can reduce bandwidth requirements while enhancing transmission efficiency and reliability of fingerprints through implementing an ``understanding-then-transmitting" strategy. 

% % The paradigm's adaptive nature allows dynamic information prioritization, eliminating redundancies and enabling flexible data exchange, which are particularly valuable for high-dimensional CSI fingerprint transmission in collaborative PLA scenarios.

Collaborative PLA utilizes multi-party cooperation to exhibit superior authentication performance. As the dimensionality of fingerprint samples increases, a key challenge is how to efficiently transmit the fingerprints estimated by collaborators to Bob. Semantic communication uses joint source-channel coding to understand the semantic content of information rather than just focusing on accurate symbol transmission, thereby reducing bandwidth requirements. It can potentially improve the efficiency and reliability of fingerprint transmission. GAI can further help enhance the performance of fingerprint transmission based on semantic communication, such as the construction of fingerprint knowledge bases and the management of fingerprint transmission models.

\subsection{Model Training}

\subsubsection{Attack Detection}
% GAI, leveraging self-supervised learning mechanisms, models the probability distribution of fingerprints without relying on manually annotated ``fingerprint-identity" mapping relationships. Although it was not originally designed for classification tasks, through architectural adjustments, it remains capable of performing unsupervised attack detection. For instance, By modifying the discriminator's functionality to output category probability distributions while distinguishing data authenticity, unsupervised classification can be achieved using GANs. Additionally, by comparing generation probabilities under different identity category hypotheses, GDM can implicitly construct detection boundaries without explicit identity labels. Furthermore, in the feature space of LLMs, representations of similar fingerprint samples naturally cluster together, while those of different users are mutually distant. With in-depth analysis of the distance or density in feature space, unsupervised attack detection can also be realized. Moreover, Meng \textit{et al.} \cite{meng2023physical} combined the advantages of AEs and VAEs to obtain superior attack detection performance when processing fine-grained, high-dimensional fingerprint samples.

GAI, leveraging self-supervised learning mechanisms, models the probability distribution of fingerprints without relying on manually annotated ``fingerprint-identity" mapping relationships. Although it was not originally designed for classification tasks, through architectural adjustments, it remains capable of performing unsupervised attack detection. For example, Meng \textit{et al.} \cite{meng2023physical} combined the advantages of AEs and VAEs to obtain superior attack detection performance when processing fine-grained, high-dimensional fingerprint samples. 
% Additionally, by modifying the discriminator's functionality to output category probability distributions while distinguishing data authenticity, unsupervised classification can be achieved using GANs. 
% Moreover, by comparing generation probabilities under different identity category hypotheses, GDM can implicitly construct detection boundaries without explicit identity labels.

\subsubsection{Multiuser Authentication}
Open-Set Recognition (OSR) provides an effective solution for authenticating multiple legitimate users simultaneously and detecting unknown attackers. OSR requires the model to not only correctly classify known classes seen during training, but also effectively recognize and reject samples of unknown classes during the testing phase.
Compared with traditional OSR methods, such as classifier-based adjustment approaches and distance metric methods, GAI has the following advantages: excelling in detecting high-dimensional unknown fingerprint samples, offering more direct detection evidence through reconstruction errors, and demonstrating stronger adaptability to unknown fingerprint categories \cite{guo2024towards}.

\subsubsection{Efficient Training and Deployment}

Channel Knowledge Map (CKM) helps predict CSI fingerprints and reduce the overhead of real-time fingerprint estimation by constructing a location-specific CSI database. 
% By building CKM, CSI-based downstream tasks such as PLA can be effectively empowered. 
Compared with traditional CKM construction methods, GAI can reduce dependence on measured CSI data, generate CSI fingerprints for specific locations, and achieve efficient transfer of different channel models.
% GANs can generate diverse fingerprint samples to meet the modeling requirements of complex environments; VAEs support probabilistic modeling and feature decoupling, making them suitable for handling incomplete fingerprint data; DPMs offer high-quality generations, suitable for detailed modeling of environment-channel relationships; and LLMs possess strong contextual understanding capabilities, supporting complex logical reasoning and knowledge transfer for cross-scenario applications. 
For instance, Fu \textit{et al.} \cite{fu2024generative} treated incomplete CKM images as conditions for training a conditional denoiser within the decoupled GDM, thereby addressing the pivotal challenge of reconstructing a comprehensive map for all relevant locations using only partially observed channel knowledge data.

\subsection{Performance Optimization}

\subsubsection{Flexibility for Eliminated Legitimate Users}

With the departure of old users, GAI-based PLA models need to forget the fingerprint characteristics of old users or incorporate the relevant distributions of new users to optimize and update authentication parameters. Four machine unlearning strategies can be applied to forget the fingerprint characteristics of old users, including gradient-based optimization, knowledge distillation, data sharding, and adding learnable layers strategies.

% Gradient-based optimization selectively erase memories associated with specific fingerprint information by meticulously adjusting model parameters. This process is achieved through the backward or forward propagation during the optimization of the loss function, aiming to effectively weaken or even eliminate learned associations while ensuring that the model's overall performance is not significantly impacted.
% Knowledge distillation establish a ``teacher-student" framework, where the model to be adjusted acts as the ``student", aiming to accurately replicate the behavior of a ``teacher" model that already possesses ideal performance.
% Data sharding divide the fingerprint dataset into multiple subsets, with each subset independently training a model. This modular design enables efficient and precise removal of target data within specific shards when needed, thereby accommodating user turnover.
% The adding learnable layers strategy endows the model with the ability to actively forget old user's fingerprint datasets by introducing additional parameters or training layers into the model. This approach avoids direct intervention in the model's inherent parameters, effectively preventing potential interference with original knowledge while enhancing the model's adaptability to changes of users.

\subsubsection{Defense Against Intelligent Attacks}
% DL models rely on high-dimensional fingerprint features for PLA. These features are inherently complex and subtly different, making PLA models susceptible to interference from minor perturbations. 
% Attackers can generate adversarial samples to deviate the fingerprint features extracted at the receiver from the true distribution, thereby deceiving PLA models \cite{papangelo2024adversarial}. Additionally, attackers can embed fingerprint data with specific trigger patterns before the training phase. The PLA model will implicitly learn the association between this pattern and legitimate identities, and activate this backdoor during the authentication phase, reducing authentication accuracy \cite{zhao2025explanation}. In response to adversarial attacks, Papangelo \textit{et al.} \cite{papangelo2024adversarial} verified that adversarial training is an effective defense method that can enhance the reliability and robustness of PLA models. Additionally, adversarial training can also defend against backdoor attacks. The generator of GANs can synthesize potential malicious fingerprint samples, while the discriminator distinguishes whether the input fingerprint data has been implanted with a backdoor or perturbed. Moreover, VAEs can identify abnormal samples through reconstruction errors. Since backdoor samples contain triggers, their reconstruction errors are significantly higher than those of normal fingerprint samples. Besides, the denoising process of GDM models can be used to eliminate fingerprint triggers, disrupting the structural consistency of the triggers.

Attackers can generate adversarial samples to deviate the fingerprint features extracted at the receiver from the true distribution, thereby deceiving PLA models. 
% Additionally, attackers can embed fingerprint data with specific trigger patterns before the training phase, and activate this backdoor during the authentication phase \cite{zhao2025explanation}. 
Papangelo \textit{et al.} \cite{papangelo2024adversarial} established an experimental platform using Software-Defined Radio (SDRs) hardware (USRP X310) and demonstrated that over 99\% of adversarial attacks can be effectively mitigated through GAN-based adversarial training.

% Additionally, adversarial training can also defend against backdoor attacks using the discriminator. 
% Moreover, VAEs can identify abnormal samples through reconstruction errors. Besides, the denoising process of GDM models can be used to eliminate fingerprint triggers, disrupting the structural consistency of the triggers.

\subsubsection{Adaptation for Different Scenarios}

GAI, especially Transformer-based LLM, with its powerful generalization ability and multi-scenario knowledge transfer, can flexibly adapt to the requirements of PLA in different application scenarios. In the pre-training phase, LLM learns fingerprint feature distributions from various scenarios. When encountering new scenarios, LLM adopts parameter-efficient fine-tuning techniques to adapt to local fingerprint features by updating some parameters, thereby avoiding the computational cost of retraining the entire model \cite{liu2025large}.
% For instance, Liu \textit{et al.} \cite{liu2025large} combined LLMs and proposed a BERT-based lightweight PLA framework to enhance the zero-trust edge IoT security.

% Parameter fine-tuning strategies can be primarily categorized into four sub-classes: additive methods, selective methods, reparameterization-based methods, and hybrid methods.

% additive methods modify the model architecture by introducing new trainable modules or parameters; selective methods enhance downstream task performance by fine-tuning only a subset of existing parameters; reparameterization-based methods transform the model architecture equivalently from one form to another via parameter conversions, involving constructing low-rank parameterizations to ensure training efficiency, while reverting to the original weight configuration during inference to maintain computational speed; and hybrid methods integrate advantages of multiple strategies or establish a unified framework by analyzing similarities between approaches.

\subsection{Lessons Learned}

As presented in Table \ref{tab2}, we compare how representative GAI models address PLA issues from each of fingerprint collection, model training, and performance optimization stages.
\begin{itemize}
\item 
During the fingerprint collection stage, GAI empowers fingerprint estimation, fingerprint augmentation, fingerprint filling, and fingerprint transmission to improve accuracy, enrich diversity, ensure completeness, and enhance transmission efficiency, respectively, thereby providing reliable fingerprint samples for training PLA models \cite{zhang2021generative,jiang2025radio,chi2024rf,zhao2024finding}.
\item 
During the model training stage, GAI-driven training strategies are tailored to specific task requirements, such as unsupervised attack detection, multi-user authentication, and efficient training and deployment, to develop PLA models \cite{meng2023physical,guo2024towards,fu2024generative}.
\item 
During the performance optimization stage, the trained PLA model undergoes further enhancement through GAI to achieve advanced capabilities, including improving flexibility for eliminated legitimate users, enhancing robust defense against intelligent attacks, and boosting adaptability across diverse scenarios \cite{papangelo2024adversarial,liu2025large}.
\end{itemize}

% Specifically, the importance of issues to PLA is first analyzed, and then the disadvantages of DAI and the advantages of GAI in solving these issues are compared. Furthermore, some existing research \cite{zhang2021generative,jiang2025radio,chi2024rf,zhao2024finding,meng2023physical,guo2024towards,fu2024generative,papangelo2024adversarial,liu2025large} is described to further illustrate the superiority of GAI.

\section{Proposed GAI-enhanced PLA Framework for Dynamic Environments}

\subsection{Challenges}
Existing CSI fingerprint-based PLA schemes face the following challenges in dynamic environments.
\begin{itemize}
\item \textbf{Rapidly Changing Fingerprint Distributions:} Frequent device movement or rapid environmental changes significantly alter the distribution characteristics of CSI fingerprints, undermining the generalization of decision boundaries learned by DAI models and leading to degraded authentication performance.
\item \textbf{Reliance on A Large Number of High-quality Fingerprints:} Most schemes exhibit heightened sensitivity to training fingerprint quantity, scale, and spatial distribution uniformity, severely restricting the generalization capability in dynamic scenarios.
\item \textbf{Channel Estimation Errors:} While time-series-based models aim to address channel variations in dynamic environments, the high-dimensional and complex structural nature of CSI fingerprints renders them highly susceptible to channel estimation errors, particularly under low SNR conditions.
\end{itemize}

\subsection{Framework Design}

\begin{figure}
\centering
\includegraphics[width=0.5\textwidth]{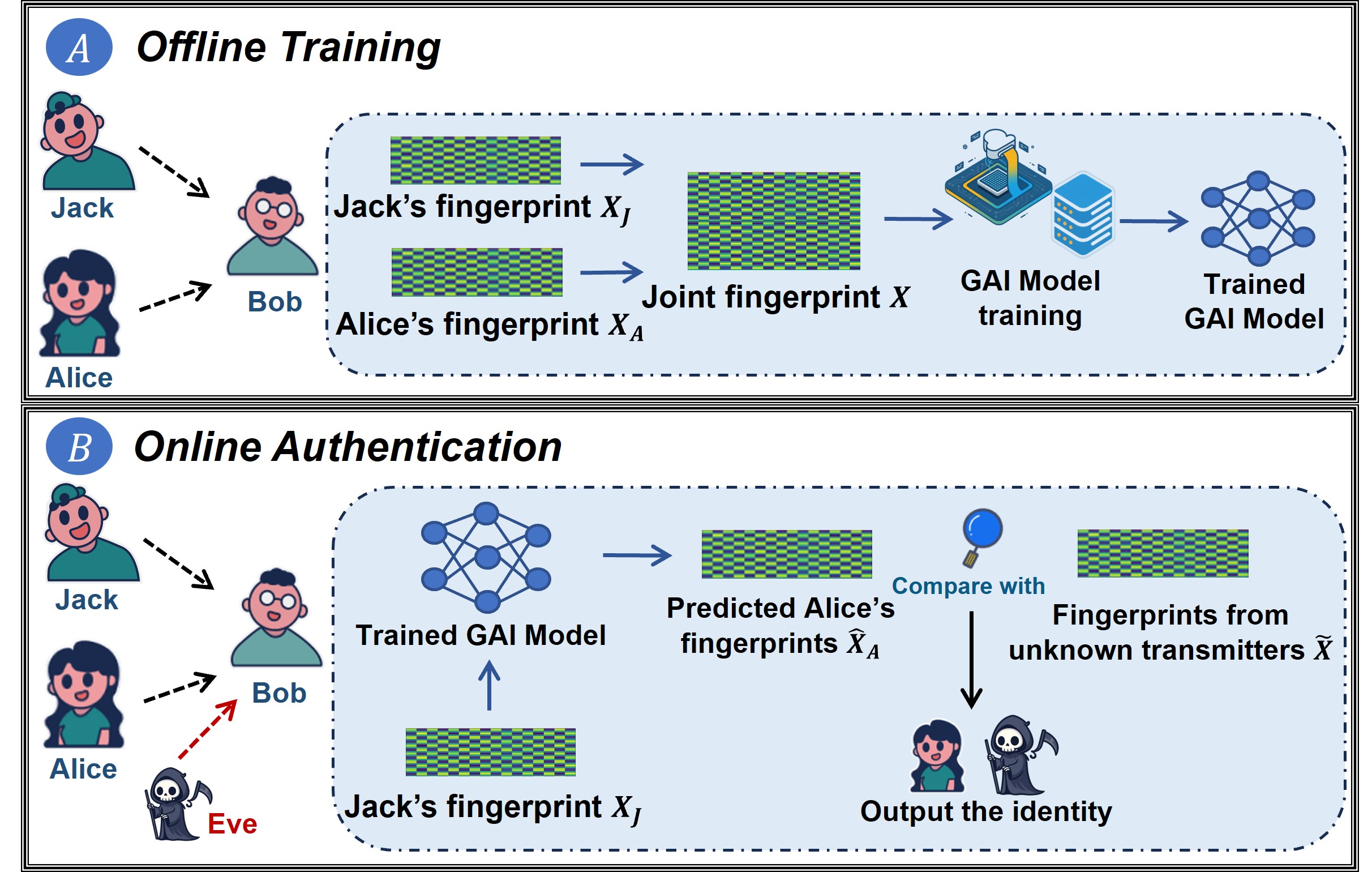}
\caption{Illustration of the proposed GAI-enhanced PLA framework for dynamic environments, where Alice is a legitimate device that requires identity authentication performed at Bob, Jack is a collaborative device, and Eve is a spoofing attacker impersonating Alice. \textit{Part (A)} denotes the offline training step including channel fingerprint pre-processing and joint distribution learning, and \textit{Part (B)} denotes the online authentication step including dynamic fingerprint prediction and identity authentication.}
\label{figure2}
\end{figure}

% We assume that Alice's signal emission frequency $f_{Alice}$ and Eve's attack frequency $f_{Eve}$ are known \cite{meng2023physical}. Therefore, we can get the proportion of Alice's fingerprint samples in the authentication stage, denoted by $\alpha = f_{Alice}/(f_{Alice}+f_{Eve})$. Then, we can obtain the detection threshold according to the difference-based sorting.
Based on channel extrapolation principles \cite{zhang2023ai}, we hypothesize correlations between the CSI fingerprints of the legitimate transmitter and collaborator, which can be learned through GAI. By leveraging Jack's fingerprints as the condition, the trained GAI model can improve the prediction accuracy of Alice's fingerprints, thus enhancing the authentication performance.
As illustrated in Figure \ref{figure2}, the proposed framework includes the following steps:
\begin{enumerate}
\item 
\textbf{Offline Training:} As illustrated in Figure \ref{figure2} \textit{Part (A)}, it includes:
\begin{enumerate}
\item \textit{Channel Fingerprint Pre-processing:} Bob obtains Alice's fingerprints $X_A$ and Jack's fingerprints $X_J$ through channel estimation. Then the fingerprints are pre-processed into one image representation for further joint analysis.
\item \textit{Joint Distribution Learning:} The joint distributions are learned through GAI model training. The problem is formulated as fitting a function $X_A=f(X_J)$ to predict $X_A$ using $X_J$.
\end{enumerate}
\item 
\textbf{Online Authentication:} As illustrated in Figure \ref{figure2} \textit{Part (B)}, it includes:
\begin{enumerate}
\item \textit{Dynamic Fingerprint Prediction:} 
Bob obtains Jack's fingerprints $X_J$, and leverages the trained GAI model to predict Alice's fingerprint $\hat{X}_A$.
\item \textit{Identity Authentication:} Bob estimates fingerprints $\tilde{X}$ from unknown transmitters, and predicts the identities by comparing the differences between $\hat{X}_A$ and $\tilde{X}$.
\end{enumerate}
\end{enumerate}

\subsection{Case Study: PLA based on GDM and Channel Extrapolation}

\begin{figure*}
\centering
\includegraphics[width=0.8\textwidth]{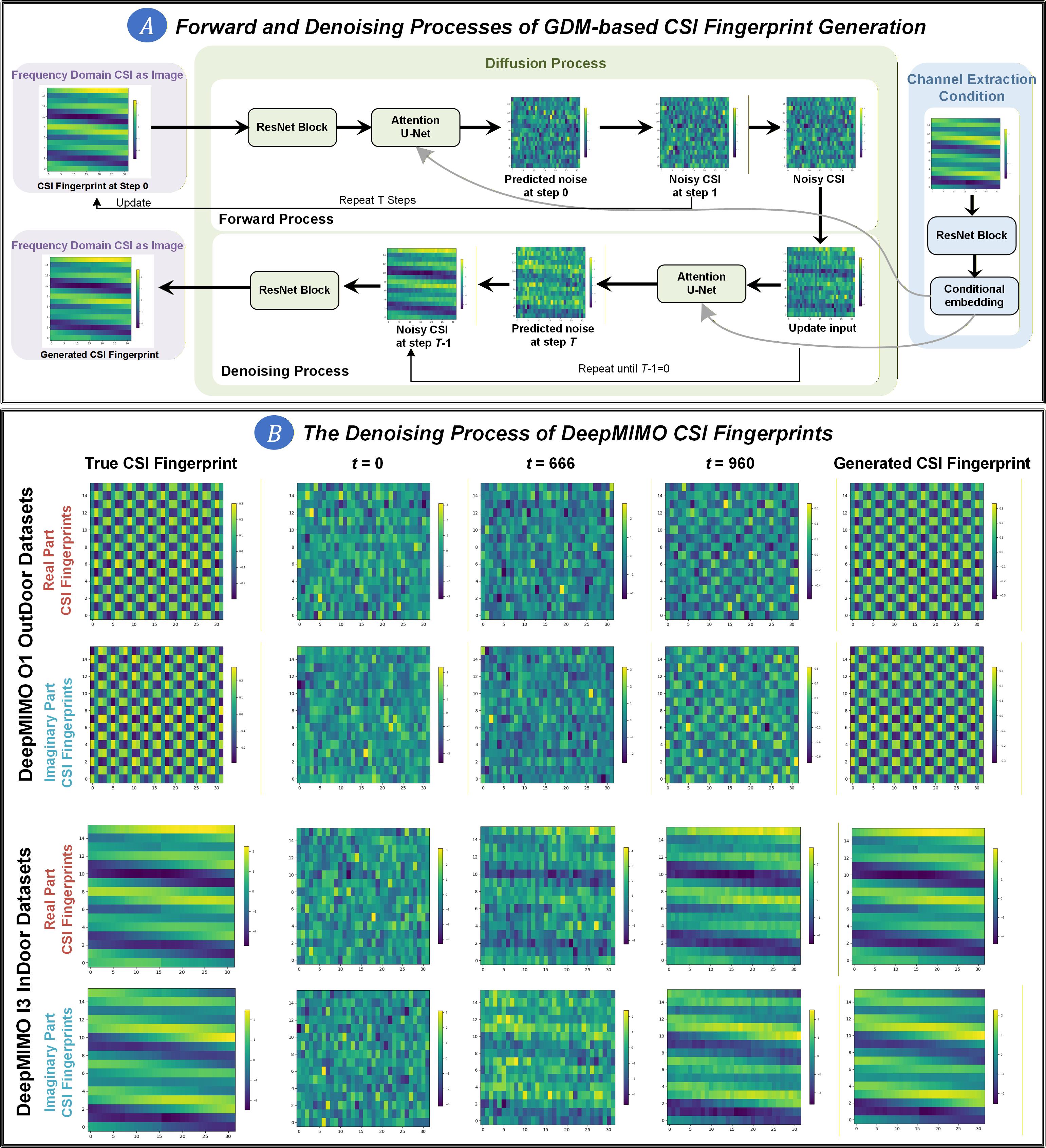}
\caption{Illustration of GDMs used for CSI fingerprint generation and prediction. To preserve the phase integrity of complex CSI fingerprints, we compute the dynamic ranges of the real and imaginary components as the differences between their maxima and minima, and then use the larger of these two ranges as a common normalization factor to linearly map both the real and imaginary values into the interval $[-1,1]$. \textit{Part (A)} denotes the training process of the GDM-based fingerprint generation, including the forward and denoising processes, and \textit{Part (B)} denotes the visual denoising process for DeepMIMO ``O1" outdoor and ``I3" indoor CSI fingerprint datasets.}
\label{figure3}
\end{figure*}

\begin{figure}
\centering
\includegraphics[width=0.5\textwidth]{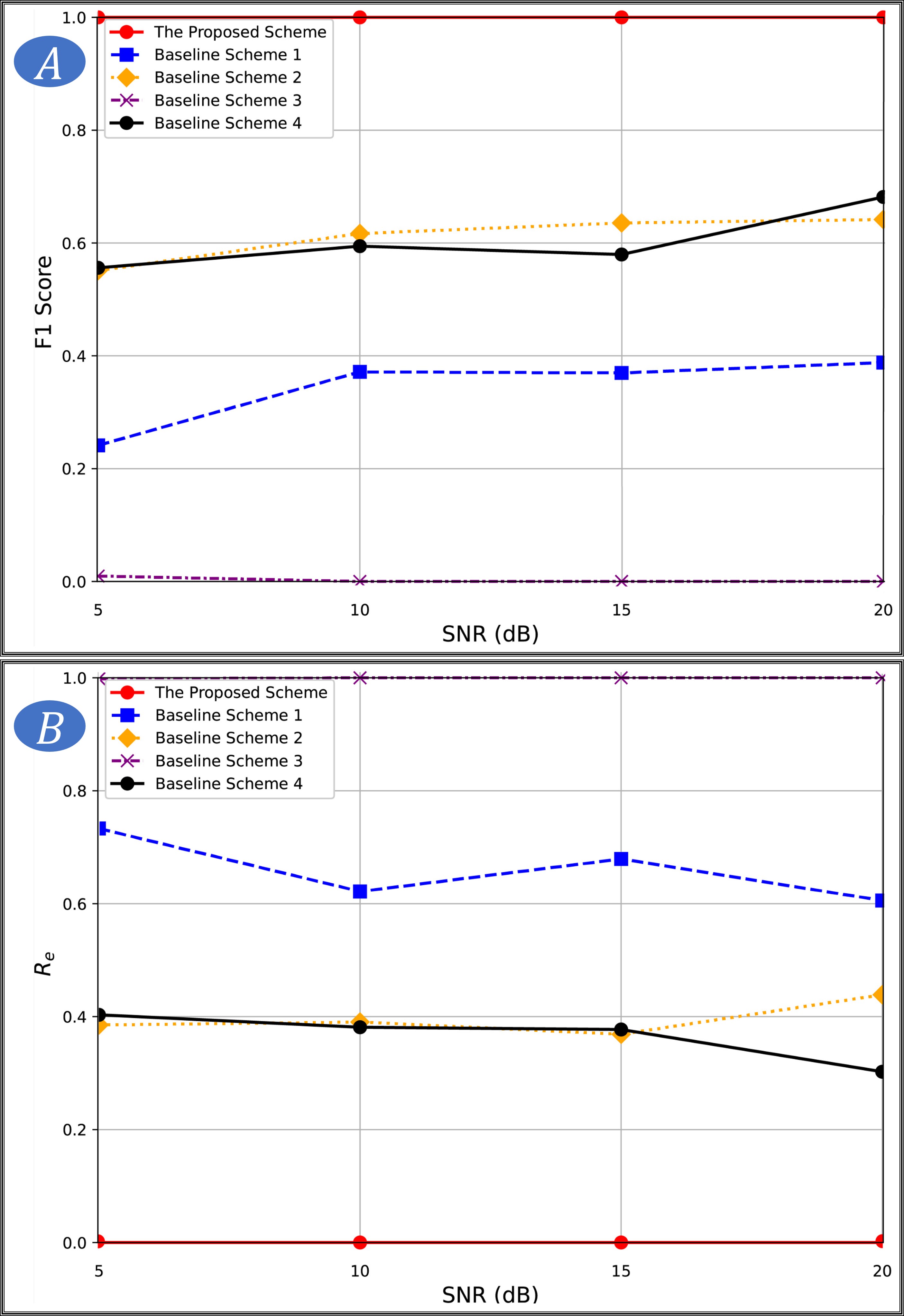}
\caption{Comparison results between the proposed GDM-based PLA scheme and four baseline schemes. \textbf{The proposed scheme} utilizes GDMs with a cross-attention mechanism to generate Alice's fingerprint at time $T$ based on Jack's fingerprint at time $T$, and compare it with the fingerprint to be authenticated at time $T$. \textbf{Baseline Scheme 1:} The VAE model with a cross-attention mechanism, is used to predict Alice's fingerprint at time $T$ and compare it with the fingerprint to be authenticated at time $T$. \textbf{Baseline Scheme 2:} The typical time series model, Long Short-Term Memory (LSTM) model, is used to predict Alice's fingerprint at time $T$ and compare it with the fingerprint to be authenticated at time $T$. \textbf{Baseline Scheme 3:} The fingerprint to be authenticated at time $T$ is compared with Jack's fingerprint at time $T$. \textbf{Baseline Scheme 4:} \textit{The Gated Recurrent Unit (GRU) model}, is used to predict Alice's fingerprint at time $T$ and compare it with the fingerprint to be authenticated at time $T$.
\textit{Part (A)} uses $F_1=(2P_{tl}P_{ta})/(P_{tl}+P_{ta}) \in[0,1]$ to evaluate the authentication performance, where $P_{tl}$ and $P_{ta}$ respectively represent correct authentication rate and false alarm rate \cite{meng2023physical}. A larger $F_1$ means better authentication performance. \textit{Part (B)} employs authentication error rate $R_e\in[0,1]$ as the metric, and a smaller $R_e$ means better authentication performance.}
\label{fig4}
\end{figure}

\subsubsection{System Model}

As shown in Figure \ref{figure2}, we assume that Alice and Jack are both stationary, while the surrounding environment changes rapidly. The fixed and close distance between Alice and Jack allows for the establishment of a stable relationship between their CSI fingerprints through channel extrapolation in the user domain \cite{zhang2023ai}.

\subsubsection{GDM Training}
Here, we train the GDM to represent $f$. As illustrated in Figure \ref{figure3} \textit{Part (A)}, it gradually adds noise to disrupt fingerprint data, and then learns to recover the original data from the noise through a reverse denoising process. The loss function of the denoising process is $\mathcal{L}
= \mathbb{E}_{X_A,X_J,\epsilon_t,t}
\left \| \epsilon_t \;-\;
\epsilon_\Theta\bigl(X_{A,t},\,t,\,X_{J}\bigr) 
\right \|_2^2 $, where $\epsilon_t$ is the Gaussian noise added to $X_{A}$ at diffusion step $t$, $X_{A,t}$ is the noisy fingerprint after $t$ diffusion steps, and $\epsilon_\Theta(X_{A,t}, t, X_J)$ is the predicted noise by model parameters $\Theta$.

\subsubsection{Complexity Analysis}
The computational complexity is denoted as $\mathcal{O}\left( T  B  L  N  \left( 2C^2 {S} + 4{S}^2 C \right)\right)$,
where $T$ is the total time step, $B$ is the batch size, $L$ is the number of layers, $N$ is the number of ResNetBlocks in each layer, and $C$ and ${S}$ denote the maximum channel number and spatial resolution, respectively. $2$ reflects the dual-branch design, while $4$ in the attention term arises from the addition of cross-attention alongside self-attention in every block.

\subsubsection{Simulation Results}

The DeepMIMO datasets are employed to generate CSI fingerprint samples.
The simulations are conducted on the computer configured as an NVIDIA RTX A6000 GPU paired with an Intel Xeon w7‑2495X CPU.

\begin{itemize}
\item \textbf{Inference Latency:} Using the Denoising Diffusion Implicit Model (DDIM) accelerator with 20 sampling steps, the total inference time is reduced to approximately 180.6 ms.
\item \textbf{Energy Consumption:} Based on hardware specifications, the system exhibits a total peak power of approximately 525 W. Consequently, the estimated energy consumption for a single authentication instance (180.6 ms) is calculated at approximately 94.8 J.
\item \textbf{Visualization of CSI Fingerprint Generation:} As illustrated in Figure \ref{figure3} \textit{Part (B)}, the frequency-domain CSI fingerprint generation in DeepMIMO ``O1" outdoor and ``I3" indoor scenarios is intuitively visualized. The comparison between the true and generated CSI fingerprints reveals that both real and imaginary fingerprint components align closely in structural distribution and amplitude range. This high-fidelity reconstruction across scenarios strongly validates GDM's accuracy and robustness in handling complex CSI fingerprint characteristics.
\item \textbf{Comparison Results:} As illustrated in Figure \ref{fig4}, the authentication performance across varying SNRs from 5 dB to 20 dB demonstrates that, the proposed scheme consistently maintains an F1 score of 1.0 and an authentication error rate of 0. This confirms its ability to accurately predict and generate CSI fingerprints for PLA. In contrast, baseline scheme 1 is constrained by limited feature fusion capabilities, while baseline schemes 2 and 4, relying solely on time-series modeling without spatial inference, fail to effectively capture complex channel mappings, resulting in significantly inferior performance. Furthermore, baseline scheme 3 fails almost completely due to the lack of an inference mechanism, validating the necessity of combining channel extrapolation with high-fidelity generative modeling to achieve highly reliable PLA.
\end{itemize}

% \subsection{Lessons Learned}
% The case study leverages channel extrapolation principles and GDMs to model and learn the joint distributions of Alice's and collaborator's channel fingerprints. Due to the excellent generation ability and noise robustness of GDMs, the authentication performance of this scheme in dynamic environment is significantly better than that of the three baseline schemes.

% In the future, the proposed scheme can be improved as follows: (1) Integrating clustering algorithms to divide the fingerprints to be authenticated into two clusters, eliminating the requirement to know the attacker's attack frequency; (2) Investigating the influence of Jack's coarse-grained fingerprints on generating Alice's fingerprints, to reduce the complexity of fingerprint estimation.

\section{Future Research Roadmap}

\subsection{Enhanced GAI Architectures for Highly reliable} Fingerprints
Existing work improves RF fingerprint generation by learning the time, frequency, and complex-valued domains \cite{chi2024rf}. To generate highly reliable fingerprints, future directions include: (1) integrating physical laws of fingerprints to ensure accuracy; (2) designing multimodal GAI models supporting cross-modal fingerprint learning in ISAC systems; and (3) combining multiple GAI models for fine-grained fingerprint processing.

\subsection{Lightweight GAI Design for Efficient PLA}
Current approaches use knowledge distillation to reduce model size and computational complexity \cite{meng2025survey}.
To enhance inference efficiency, future work should: (1) optimize GAI frameworks through efficient sampling techniques tailored for GDMs; (2) integrate quantization to implement low-precision parameters, reducing storage and computational demands;
and (3) develop plug-and-play hierarchical modules enabling adaptive component selection during deployment based on resource constraints.

\subsection{Generalization Enhancement for GAI-based PLA}
Existing methods employ data augmentation to increase fingerprint \cite{chi2024rf} or combine pre-training and fine-tuning to improve the generalization of LLM-based PLA \cite{liu2025large}.
Future research should: (1) design conditional GAI models for environment-specific fingerprint generation; (2) introduce meta-learning or incremental learning to adapt GAI models to environmental changes; (3) adopt domain adaptation to learn fingerprint distribution gaps between training and target environments; and (4) integrate digital twins to optimize the generalization of GAI models in unseen environments.

\subsection{Securing GAI-based PLA}

Adversarial training mitigates adversarial attacks \cite{papangelo2024adversarial}, but some intelligent attacks such as backdoor and poisoning attacks oriented to GAI models remain unresolved. Future directions include: (1) deploying adversarial cleaning, model pruning, robust optimization, and trusted execution environments (TEEs) to defend against backdoor and poisoning attacks across pre-processing, training, and authentication stages; (2) using machine unlearning to erase specific knowledge and counter adversarial attacks; (3) applying differential privacy, knowledge distillation, and regularization to address membership inference attacks and protect the privacy of training fingerprint samples; and (4) implementing federated learning or secure multi-party computation for distributed training \cite{hou2025split}, avoiding centralized PLA limitations.

\subsection{Promoting the Industrialization of GAI-based PLA}

GAI-based PLA solutions for IoT \cite{zeng2022adaptive}, IIoT \cite{meng2023physical}, and 6G edge intelligence \cite{liu2025large} are developed. Industrialization requires:
(1) establishing benchmark fingerprint datasets and GAI models for representative scenarios;
(2) designing interdisciplinary PLA methods to leverage GAI technologies; and
(3) establishing diversified evaluation standards to comprehensively assess the real-world deployment effectiveness of GAI-based PLA.

\section{Conclusions}

In this paper, we first provide a concise overview of identity authentication techniques and GAI models. We then analyze the limitations of applying DAI in PLA and explore how GAI can address these challenges. Furthermore, we propose a structured taxonomy for GAI-enhanced PLA methods, delineating three key stages of the PLA pipeline: fingerprint collection, model training, and performance optimization. Specifically, GAI contributes to (1) fingerprint collection: enhancing accuracy, enriching diversity, ensuring completeness, and improving transmission efficiency of collected fingerprints; (2) model training: tailoring models to meet specific task requirements; and (3) performance optimization: increasing flexibility to accommodate legitimate users, strengthening robustness against intelligent adversarial attacks, and boosting adaptability across diverse operational scenarios. Additionally, we introduce a novel PLA framework leveraging channel extrapolation and GAI, supported by a case study using GDMs to demonstrate GAI’s superiority in enhancing PLA robustness. Finally, we outline a forward-looking research roadmap for GAI-based PLA.

\bibliography{ref.bib}
\bibliographystyle{IEEEtran}

\vfill

\end{document}